\documentclass[aps,prd,preprint,showkeys,showpacs,superscriptaddress]{revtex4-1}

\usepackage{latexsym}
\usepackage{amsmath,amssymb}
\usepackage{graphicx}
\usepackage{subfigure}
\usepackage{xcolor}
\usepackage{hyperref}     
\hypersetup{colorlinks,%
  citecolor=blue,%
  linkcolor=cyan,%
 pdftex}

\begin{document}


\title{Superradiance and statistical entropy of hairy black
  hole in three dimensions}

\author{Myungseok Eune} %
\email[]{younms@sogang.ac.kr} %
\affiliation{Research Institute for Basic Science, Sogang University,
  Seoul, 121-742, Republic of Korea} %

\author{Yongwan Gim} %
\email[]{yongwan89@sogang.ac.kr} %
\affiliation{Department of Physics, Sogang University, Seoul 121-742,
  Republic of Korea} %

\author{Wontae Kim} %
\email[]{wtkim@sogang.ac.kr} %
\affiliation{Research Institute for Basic Science, Sogang University,
  Seoul, 121-742, Republic of Korea} %
\affiliation{Department of Physics, Sogang University, Seoul 121-742,
  Republic of Korea} %
\affiliation{Center for Quantum Spacetime, Sogang University, Seoul
  121-742, Republic of Korea} %

\date{\today}

\begin{abstract}
  We calculate the statistical entropy of a rotating hairy black hole
  by taking into account superradiant modes in the brick wall
  method. The UV cutoff is independent of the gravitational hair,
  which gives the well-defined area law of the entropy.  It can be
  shown that the angular momentum and the energy of matter field
  depend on the gravitational hair.  For the vanishing gravitational
  hair, it turns out that the energy for matter is related to both the
  black hole mass and the black hole angular momentum whereas the
  angular momentum for matter field is directly proportional to the
  angular momentum of the black hole.
\end{abstract}


\maketitle

\section{Introduction}
\label{sec:intro}

There has been much attention to three-dimensional topological massive
gravity~\cite{Deser:1981wh,Deser:1982vy} because it has rich
structures even though it is lower dimensional gravity. Recently, new
massive gravity~\cite{Bergshoeff:2009hq, Bergshoeff:2009aq} has been
also intensively studied, in particular, it can be shown that there is
a new type of a rotating black hole solution apart from the rotating
Banados-Teitelboim-Zanelli (BTZ) black hole~\cite{Banados:1992wn}. A
new rotating black hole has three hairs: two of them are mass and
angular momentum and the other corresponds to the gravitational
hair~\cite{Oliva:2009ip, Giribet:2009qz, Kwon:2011jz, Perez:2011qp}.

On the other hand, the statistical origin of the entropy for black holes
has been studied in terms of the brick
wall method \cite{tHooft:1984re}. Subsequently, there have been extensive
applications of the brick wall method to various black
holes~\cite{Mann:1990fk, Ghosh:1994wb, Kay:2011np}. 
In connection with the rotating hairy black hole,
there exists a superradiant mode so that the
brick wall method is nontrivial, and it should be treated
carefully~\cite{Ho:1998du}. Moreover, a thin-layer method has been
introduced~\cite{Liu:2001qj, Zhou:2004yq}, where it is in thermal
equilibrium locally and the divergent term due to a box with infinite
size does not appear anymore. For a thin layer near the horizon, this
method is valid if the proper thickness is taken keeping the local
equilibrium state, since the degree of freedom is dominant near the
horizon.

In this paper, we would like to calculate the entropy of the rotating
hairy black hole \cite{Oliva:2009ip} from new massive gravity
\cite{Bergshoeff:2009hq, Bergshoeff:2009aq} by using the brick wall
method \cite{tHooft:1984re} in thin-layer approximations
\cite{Liu:2001qj, Zhou:2004yq}, where the angular speed of a particle
near the horizon can be approximately fixed to a constant.  So we will
show how to take into account the superradiant mode in the entropy
calculation of the rotating hairy black hole which has complicated
metric components. The entropy of the hairy black hole in connection
with the superradiant mode for the rotating case deserves to be
studied.  In Sec.~\ref{sec:hairyBH}, a new hairy rotating black hole
and its thermodynamic quantities are introduced. Additionally, one can
write down the explicit form of the horizons and the radius of the
ergosphere analytically.  In Sec.~\ref{sec:thermo}, we calculate the
thermodynamic quantities such as the entropy, the angular momentum,
and the energy by considering the superradiant and the nonsuperradiant
modes simultaneously.  A summary is given in Sec.~\ref{sec:discus}.

\section{Rotating Hairy Black Hole}
\label{sec:hairyBH}

We consider the rotating black hole described
by~\cite{Oliva:2009ip, Giribet:2009qz, Kwon:2011jz, Perez:2011qp}
\begin{align}
  \label{metric}
  ds^2 = -N^2 f dt^2 + \frac{dr^2}{f} + r^2 (d\phi - \Omega_0 dt)^2,
\end{align}
where $N(r) = 1 + b \ell^2 (1 - \eta) / (4\Sigma)$, $f(r)=
(\Sigma^2/r^2) \left[\Sigma^2/\ell^2 + b(1 + \eta)\Sigma /2 + b^2
  \ell^2 (1 - \eta)^2 /16 - \mu \eta \right]$, $ \Omega_0 (r) = a (\mu
- b\Sigma) / (2r^2)$, and $ \Sigma(r)^2 = r^2 - \mu \ell^2(1 - \eta)
/2 - b^2 \ell^4 (1 - \eta)^2 /16 $.  $a$, $b$, and $\mu$ are
integration constants and $\Lambda$ and $\eta$ have been defined as
$\Lambda = -1/(2\ell^2)$ and $\eta \equiv \sqrt{1 - a^2 /
  \ell^2}$. For the limit of $b=0$, the metric functions reduce to
$N=1$, $f = -\mu + r^2/\ell^2 +\mu^2 a^2/(4r^2)$, and $\Omega_0 = \mu
a/(2r^2)$, and Eq.~\eqref{metric} simply describes the rotating BTZ
black hole \cite{Banados:1992wn}.  Note that the range of the rotating
parameter $a$ is given by $-\ell \le a \le \ell$.  Then the
Arnowitt-Deser-Misner (ADM) mass and the ADM angular momentum can be
obtained as~\cite{Kwon:2011jz}
\begin{align}
  \mathcal{M}= \frac{\mu}{4G} + \frac{b^2 \ell^2}{16G}, \qquad
  \mathcal{J} = \mathcal{M} a,
\end{align}
respectively. $G$ is a Newton constant in three dimensions, which is
set to $G=1$ for convenience.  The entropy has been also obtained as
$\mathcal{S} = \pi \ell \sqrt{2 \mathcal{M}
  (1+\eta)}$~\cite{Oliva:2009ip, Giribet:2009qz, Perez:2011qp}. Now,
the condition of $f(r) = 0$ gives
\begin{align}
  r_\pm &= \ell \sqrt{2(1 + \eta)} \left(\sqrt{\mathcal{M}} \pm
    \frac{|b|\ell}{4} \sqrt{\eta} \right), \label{r:pm} \\
  r_0 &= \ell \sqrt{2(1 - \eta)} \left[\mathcal{M} - \frac{b^2
        \ell^2}{32} (1 + \eta) \right]^{1/2}, \label{r:0}
\end{align}
which yields the horizon $r_+$.  Equation~\eqref{metric} describes the
rotating BTZ black hole, which has two horizons of $r_\pm = \ell
\sqrt{\mu(1 + \eta)/2}$ and $r_0 = \ell \sqrt{\mu (1 -
  \eta)/2}$. Especially for $a=0$, the black hole has two horizons
$r_\pm = 2\ell (\sqrt{\mathcal{M}} \pm |b|\ell/4)$ with $r_0 = 0$.
Next, the Hawking temperature of the black hole can be obtained from
the surface gravity,
\begin{align}
T_{\rm H} &= \frac{1}{4\pi}N(r_+) f'(r_+)  \nonumber \\
&= \frac{\eta}{\pi\ell} \sqrt{\frac{2\mathcal{M}}{1+\eta}}. \label{TH}
\end{align}
Consequently, the
first law of thermodynamics $d\mathcal{M} = T_{\rm H} d\mathcal{S} +
\Omega_H d\mathcal{J}$ is satisfied.

The maximum $\Omega_+$ and the minimum $\Omega_-$ of the angular
velocity for a particle are given by
\begin{align}
  \Omega_\pm (r) &= \Omega_0 \pm
  \frac{N}{r}\sqrt{f}. \label{def:Omega:pm}
\end{align}
Note that the angular velocity of a particle on the event horizon
becomes
\begin{align}
  \Omega_H = \Omega_0 (r_+) = \frac{1}{\ell}
  \sqrt{\frac{1-\eta}{1+\eta}}, \label{Omega:H}
\end{align}
and the radius $r_e$ of the ergosphere is explicitly
written as
\begin{align}
  r_e &= 2 \ell \sqrt{\mathcal{M} + \frac{b^2 \ell^2}{16} + \frac{|b|
      \ell}{4} \sqrt{2(1 + \eta)} \left[\mathcal{M} + \frac{b^2
          \ell^2}{32} (1-\eta) \right]^{1/2}}, \label{r:e}
\end{align}
which can be calculated from $\Omega_- (r_e) = 0$.

\section{Statistical entropy}
\label{sec:thermo}

Now, we consider a scalar field in a thin layer between $r_+ + h$ to
$r_+ + h + \delta$ with $h \ll r_+$ and $\delta \ll r_+$, where $h$ is
a cutoff parameter and $\delta$ is a small constant related to the
thickness of the thin layer. It satisfies the massless Klein-Gordon equation,
$ \square \Phi(t,r,\phi)=0$.  Assuming
$\Phi(t,r,\phi)=\Psi_{\omega m}(r)~e^{-i \omega t+i m \phi}$, we
obtain $ r N \partial_r (rNf \partial_r \Psi_{\omega m}) + r^2 N^2 f^2
k^2 \Psi_{\omega m} = 0$, where $ k(r;\omega,m) = N^{-1} f^{-1}
\sqrt{(\omega - \Omega_+ m)(\omega - \Omega_- m)}$.  In the WKB
approximation with $\Psi \sim e^{i S(r)}$, $k$ is the radial momentum
defined by $k = \partial S/\partial r$. Therefore, the number of
states less than the energy $\omega$ and the angular momentum $m$ is
given by 
\begin{align}
  n(\omega,m) = \frac{1}{\pi} \int^{r_h+h+\delta}_{r_h+h} dr ``k"(r;
  \omega, m), \label{def:n}
\end{align}
where$``k"(r; \omega, m) = k(r; \omega, m)$ if $k^2 > 0$ and $``k"(r;
\omega, m) = 0$ if $k^2 < 0$. The free energy of a rotating black hole
should be written as~\cite{Ho:1998du}
\begin{align}
  F=F_{\mathrm{NS}} + F_{\mathrm{SR}}, \label{F:sum:NS:SR}
\end{align}
where 
\begin{align} 
  \beta F_{\rm NS} &= \sum_{\lambda \notin \mathrm{SR}} \int d\omega
  g(\omega,m) \ln[1 - e^{-\beta(\omega - m\Omega_H)}], \label{F:NS:def} \\
  \beta F_{\rm SR} &= \sum_{\lambda \in \mathrm{SR}} \int d\omega
  g(\omega,m) \ln[1 - e^{\beta(\omega - m\Omega_H)}], \label{F:SR:def}
\end{align}
where the ``NS'' and ``SR'' denote the nonsuperradiant mode with
$\omega - m \Omega_H >0$ and superradiant mode with $\omega - m
\Omega_H < 0$, respectively, $\lambda$ is the set of $(\omega, m)$,
and the density of the number of states is given by $g(\omega,m) =
dn/d\omega$ for the NS mode and $g(\omega,m) = - dn/d\omega$ for the
SR mode.  Substituting Eq.~\eqref{def:n} into Eqs.~\eqref{F:NS:def}
and \eqref{F:SR:def}, we obtain
\begin{align} 
  \beta F_{\mathrm{NS}} = &- \frac{\beta}{\pi} \int dr \sum_m \int
  d\omega \frac{``k"(r; \omega, m)}{e^{\beta(\omega - \Omega_H m)} - 1} \notag\\
  &+ \frac{1}{\pi} \int dr \sum_m \left. ``k"(r; \omega, m)\ln[1 -
    e^{-\beta(\omega - \Omega_H m)}]
  \right|^{\omega_{\rm max}(m)}_{\omega_{\rm min}(m)}, \label{F:NS:n} \\
  \beta F_{\mathrm{SR}} = &-\frac{\beta}{\pi} \int dr \sum_m \int
  d\omega
  \frac{``k"(r; \omega, m)}{e^{-\beta(\omega - \Omega_H m)} - 1} \notag\\
  &- \frac{1}{\pi} \int dr \sum_m \left.  ``k"(r; \omega, m) \ln[1 -
    e^{\beta(\omega - \Omega_H m)}] \right|^{\omega_{\rm
      max}(m)}_{\omega_{\rm min}(m)}, \label{F:SR:n}
\end{align}
where $\omega_{\rm max}(m)$ and $\omega_{\rm min}(m)$ denote the
maximum and the minimum of $\omega$ for a given $m$ in each mode,
respectively.  For convenience, Eq.~\eqref{F:NS:n} can be rewritten as
\begin{align}
  F_{\rm NS} \equiv F_{\rm NS}^{(m>0)} + F_{\rm NS}^{(m<0)},
\end{align}
where
\begin{align} 
  \beta F_{\mathrm{NS}}^{(m>0)} &= -\frac{\beta}{\pi} \int^{r_+ + h +
    \delta}_{r_+ + h} \frac{dr}{Nf} \int^\infty_0 dm
  \int^\infty_{\Omega_+ m} d\omega \frac{\sqrt{(\omega - \Omega_+
      m)(\omega - \Omega_- m)}}{e^{\beta(\omega - \Omega_H
      m)} - 1}, \label{F:NS:m:positive} \\
  \beta F_{\mathrm{NS}}^{(m<0)} &= -\frac{\beta}{\pi} \int^{r_+ + h +
    \delta}_{r_+ + h} \frac{dr}{N f} \int^0_{-\infty} dm \int^\infty_0
  d\omega \frac{\sqrt{(\omega - \Omega_+ m)(\omega - \Omega_-
      m)}}{e^{\beta(\omega - \Omega_H m)} - 1} \notag\\
  &\quad -\frac{1}{\pi} \int^{r_+ + h + \delta}_{r_+ + h} \frac{dr}{N
    f} \int^0_{-\infty} dm \sqrt{\Omega_+ \Omega_- m^2}\ln \left(1 -
  e^{\beta \Omega_H m} \right). \label{F:NS:m:negative}
\end{align}
From Eq.~\eqref{F:SR:n}, the free energy of the SR mode is written as
\begin{align} 
  \beta F_{\mathrm{SR}} =& -\frac{\beta}{\pi} \int^{r_+ + h +
    \delta}_{r_+ + h} \frac{dr}{N f} \int^\infty_0 dm \int^{\Omega_-
    m}_0 d\omega \frac{\sqrt{(\omega - \Omega_+ m)(\omega - \Omega_-
      m)}}{e^{-\beta(\omega - \Omega_H m)} - 1} \notag\\
  &+ \frac{1}{\pi} \int^{r_+ + h + \delta}_{r_+ + h} \frac{dr}{N f}
  \int^{\infty}_0 dm \sqrt{\Omega_+ \Omega_- m^2}\ln \left(1 -
    e^{-\beta \Omega_H m} \right). \label{F:SR:m}
\end{align}
Then, the total free energy which consists of the nonsuperradiant and
superradiant modes~\eqref{F:sum:NS:SR} becomes
\begin{align} 
  F &= -\frac{\zeta(3)}{4\beta^{3}} \int^{r_+ + h + \delta}_{r_+ + h}
  \frac{dr}{Nf} \frac{(\Omega_+ - \Omega_-)^2}{(\Omega_+ -
    \Omega_H)^{3/2}(\Omega_{H} - \Omega_-)^{3/2}}, \label{F:total:dr}
\end{align}
which leads to
\begin{align} 
  F &= - \frac{\zeta(3)}{\beta^{3}} \frac{2r_+}{N(r_+)^2
    f'(r_+)^{3/2}} \left(\frac{1}{\sqrt{h}} - \frac{1}{\sqrt{h +
        \delta}}\right), \label{F:beta}
\end{align}
in the leading order of the cutoff and the thickness.  Note that the
second term in the free energy for the positive mode in
Eq.~\eqref{F:NS:m:negative} and the second term in the free energy for
the superradiant mode in~\eqref{F:SR:m} canceled out.  Thus the
entropy can be simplified as
\begin{align}
  S &= \left. \beta^2 \frac{\partial F}{\partial \beta} \right|_{\beta = \beta_{\rm H}} \notag \\
  &= \frac{3 \zeta(3)}{8 \pi^2} r_+ \sqrt{f'(r_+)}
  \left(\frac{1}{\sqrt{h}} - \frac{1}{\sqrt{h + \delta}}
  \right), \label{S:h}
\end{align}
where $\beta_{\rm H}$ is defined as the inverse of the Hawking
temperature $T_{\rm H}$.  The proper lengths for the UV cutoff
parameter and the thickness are defined by $\bar{h} \equiv \int^{r_+ +
  h}_{r_+} dr \sqrt{g_{rr}} \simeq 2 \sqrt{h} / \sqrt{f'(r_+)}$ and
$\bar{\delta} \equiv \int^{r_+ + h + \delta}_{r_+ + h} dr
\sqrt{g_{rr}} \simeq 2 (\sqrt{h + \delta} - \sqrt{h}) /
\sqrt{f'(r_+)}$. Then, the entropy is written as $S = 3 \zeta(3) r_+
\bar{\delta} / [4\pi^2 \bar{h} (\bar{\delta} + \bar{h})]$. Recovering
dimensions, the entropy becomes
\begin{align}
  S=\frac{c^3 A}{4G\hbar} \frac{3\zeta(3) \ell_{\rm
      P}\bar\delta}{2\pi^3 \bar{h} (\bar{h} +
    \bar\delta)}, \label{S:dim}
\end{align}
where $A \equiv 2\pi r_+$ and $\ell_{\rm P} \equiv \hbar G/c^3$ are
the area of the event horizon and the three-dimensional Plank length,
respectively. If the cutoff is chosen as $\bar{h}(\bar{h} +
\bar\delta)/\bar\delta = [3\zeta(3) /(2\pi^3)] \ell_{\rm P}$, the
entropy~\eqref{S:dim} agrees with the Bekenstein-Hawking entropy
$S_{\rm BH} = c^3 A/(4G \hbar)$.

Finally, let us calculate angular momentum of matter, which becomes
\begin{align}
  J &= \left. -\frac{\partial F}{\partial
      \Omega_{\mathrm{H}}} \right|_{\beta=\beta_{\mathrm{H}}}
  \notag \\
  &= \frac{a}{2} \left(\sqrt{\mathcal{M}} +  \frac{|b| \ell}{4} \sqrt{\eta}
  \right) \left[\sqrt{\mathcal{M}}+ \frac\ell8  (b + |b|)
    \left(\sqrt\eta + \frac{1}{\sqrt\eta}\right) \right], \label{J:h}
\end{align}
and the internal energy of the system is written as
\begin{align}
  E &= F_\mathrm{H} + \beta^{-1}_{\mathrm{H}} S+
  \Omega_\mathrm{H} J \notag \\
  &= \frac{1}{6} \left(\sqrt{\mathcal{M}}+ \frac{|b| \ell}{4}
    \sqrt{\eta} \right) \left[(3+ \eta)\sqrt{\mathcal{M}} + \frac{3
      \ell}{8\sqrt\eta} (b + |b|)(1-\eta^2) \right]. \label{E:h}
\end{align}
Note that the angular momentum \eqref{J:h} and the energy \eqref{E:h}
of matter have well-defined limits and they are compatible with the
results in Ref.~\cite{Ho:1998du} for $b=0$.

On the other hand, it would be interesting to note that a partition
function from free energy~\eqref{F:beta} can be compared with the
result for the partition function of the corresponding two-dimensional
conformal field theory (CFT) on the boundary of three-dimensional
anti-de Sitter (AdS) spacetime~\cite{Hawking:1998kw}. For this
purpose, we write down the free energy~\eqref{F:beta} by using
Eqs.~\eqref{TH} and~\eqref{Omega:H} along with the proper lengths
$\bar{h}$ and $\bar\delta$ as
\begin{align}
  F &= -\frac{\zeta(3)}{2\pi (\beta_{H}/\ell)^2 [1 - (\ell
    \Omega_{H})^2]} \frac{\bar{\delta}}{\bar{h} (\bar{\delta} +
    \bar{h})}, \label{F:beta:H}
\end{align}
where we restricted to the case of $b=0$ and identified $\beta$ with
$\beta_{\rm H}$. If one chooses the cutoff as \(\bar{h}(\bar{\delta} +
\bar{h}) / \bar{\delta} = [3\zeta(3) / \pi^3] \ell_{\rm P}\), then
the free energy~\eqref{F:beta:H} is simplified as
\begin{align}
  F = -\frac{\pi^2}{6(\beta_{H}/\ell)^2 [1 - (\ell \Omega_{H})^2]
    \ell_{\rm P}}. \label{F:beta:cutoff}
\end{align}
In order to write down the free energy in terms of dimensionless
quantities, we rescale the free energy, the inverse Hawking
temperature, and the angular velocity at the horizon by \(\ell_{\rm P}
F \to F\), \(\beta_{\rm H}/\ell \to \beta\), and \(\ell \Omega_{\rm H}
\to \Omega\), respectively. Then, the free energy becomes \(F =
-\pi^2/[6\beta^2 (1 - \Omega^2)]\). Since the relation between the
partition function $Z$ and the free energy is given by \(\beta F = -
\ln Z\), we can obtain
\begin{align}
  \ln Z = \frac{\pi^2}{6\beta (1- \Omega^2)}, \label{partition}
\end{align}
which agrees with the result given in Ref.~\cite{Hawking:1998kw}.
However, it may depend on the cutoff within our brick wall formulation
so that the coefficient can be adjusted. As a result, the degrees of
freedom near the horizon can be described by the boundary degrees of
freedom. In fact, the bulk degrees of freedom can be read off from the
boundary degrees of freedom from the AdS/CFT while the bulk degrees of
freedom can be also described by the degrees of freedom near the
horizon based on the brick wall formalism. Combining these two
notions, the boundary degrees of freedom at both ends can be
connected.


\section{Summary}
\label{sec:discus}

In the course of calculations, the second term of the free energy for
the positive mode in Eq.~\eqref{F:NS:m:negative} and the second term
of the free energy for the superradiant mode in~\eqref{F:SR:m}
canceled out so that from the simplified resulting free energy we have
obtained the statistical entropy satisfying the area law by
determining the UV cutoff which is independent of the hairs of the
black hole, and additionally derived the angular momentum and the
energy of matter field.

The energy $E$ is always positive and it depends on the mass of the
black hole, the angular momentum of the black hole, and the gravitational
hair $b$.  For the limit of $b=0$, the angular momentum can be reduced
to $J=\frac12\mathcal{J}$ and $E= \frac12 \mathcal{M}+ \frac16
\sqrt{\mathcal{M}^2-\mathcal{J}^2/\ell^2}$. It means that the angular
momentum of the matter is directly proportional to that of the black
hole while the energy is related to the mass and angular momentum of
the black hole simultaneously.

\begin{acknowledgments}
  This work was supported by the Sogang University Research Grant
  201310022 (2013).
\end{acknowledgments}


\bibliographystyle{apsrev4-1} 
\bibliography{references}

\begin{thebibliography}{17}%
\makeatletter
\providecommand \@ifxundefined [1]{%
 \@ifx{#1\undefined}
}%
\providecommand \@ifnum [1]{%
 \ifnum #1\expandafter \@firstoftwo
 \else \expandafter \@secondoftwo
 \fi
}%
\providecommand \@ifx [1]{%
 \ifx #1\expandafter \@firstoftwo
 \else \expandafter \@secondoftwo
 \fi
}%
\providecommand \natexlab [1]{#1}%
\providecommand \enquote  [1]{``#1''}%
\providecommand \bibnamefont  [1]{#1}%
\providecommand \bibfnamefont [1]{#1}%
\providecommand \citenamefont [1]{#1}%
\providecommand \href@noop [0]{\@secondoftwo}%
\providecommand \href [0]{\begingroup \@sanitize@url \@href}%
\providecommand \@href[1]{\@@startlink{#1}\@@href}%
\providecommand \@@href[1]{\endgroup#1\@@endlink}%
\providecommand \@sanitize@url [0]{\catcode `\\12\catcode `\$12\catcode
  `\&12\catcode `\#12\catcode `\^12\catcode `\_12\catcode `\%12\relax}%
\providecommand \@@startlink[1]{}%
\providecommand \@@endlink[0]{}%
\providecommand \url  [0]{\begingroup\@sanitize@url \@url }%
\providecommand \@url [1]{\endgroup\@href {#1}{\urlprefix }}%
\providecommand \urlprefix  [0]{URL }%
\providecommand \Eprint [0]{\href }%
\providecommand \doibase [0]{http://dx.doi.org/}%
\providecommand \selectlanguage [0]{\@gobble}%
\providecommand \bibinfo  [0]{\@secondoftwo}%
\providecommand \bibfield  [0]{\@secondoftwo}%
\providecommand \translation [1]{[#1]}%
\providecommand \BibitemOpen [0]{}%
\providecommand \bibitemStop [0]{}%
\providecommand \bibitemNoStop [0]{.\EOS\space}%
\providecommand \EOS [0]{\spacefactor3000\relax}%
\providecommand \BibitemShut  [1]{\csname bibitem#1\endcsname}%
\let\auto@bib@innerbib\@empty
\bibitem [{\citenamefont {Deser}\ \emph
  {et~al.}(1982{\natexlab{a}})\citenamefont {Deser}, \citenamefont {Jackiw},\
  and\ \citenamefont {Templeton}}]{Deser:1981wh}%
  \BibitemOpen
  \bibfield  {author} {\bibinfo {author} {\bibfnamefont {S.}~\bibnamefont
  {Deser}}, \bibinfo {author} {\bibfnamefont {R.}~\bibnamefont {Jackiw}}, \
  and\ \bibinfo {author} {\bibfnamefont {S.}~\bibnamefont {Templeton}},\ }\href
  {\doibase 10.1016/0003-4916(82)90164-6} {\bibfield  {journal} {\bibinfo
  {journal} {Ann. Phys. (N.Y.)}\ }\textbf {\bibinfo {volume} {140}},\ \bibinfo
  {pages} {372} (\bibinfo {year} {1982}{\natexlab{a}})}\BibitemShut {NoStop}%
\bibitem [{\citenamefont {Deser}\ \emph
  {et~al.}(1982{\natexlab{b}})\citenamefont {Deser}, \citenamefont {Jackiw},\
  and\ \citenamefont {Templeton}}]{Deser:1982vy}%
  \BibitemOpen
  \bibfield  {author} {\bibinfo {author} {\bibfnamefont {S.}~\bibnamefont
  {Deser}}, \bibinfo {author} {\bibfnamefont {R.}~\bibnamefont {Jackiw}}, \
  and\ \bibinfo {author} {\bibfnamefont {S.}~\bibnamefont {Templeton}},\ }\href
  {\doibase 10.1103/PhysRevLett.48.975} {\bibfield  {journal} {\bibinfo
  {journal} {Phys. Rev. Lett.}\ }\textbf {\bibinfo {volume} {48}},\ \bibinfo
  {pages} {975} (\bibinfo {year} {1982}{\natexlab{b}})}\BibitemShut {NoStop}%
\bibitem [{\citenamefont {Bergshoeff}\ \emph
  {et~al.}(2009{\natexlab{a}})\citenamefont {Bergshoeff}, \citenamefont
  {Hohm},\ and\ \citenamefont {Townsend}}]{Bergshoeff:2009hq}%
  \BibitemOpen
  \bibfield  {author} {\bibinfo {author} {\bibfnamefont {E.~A.}\ \bibnamefont
  {Bergshoeff}}, \bibinfo {author} {\bibfnamefont {O.}~\bibnamefont {Hohm}}, \
  and\ \bibinfo {author} {\bibfnamefont {P.~K.}\ \bibnamefont {Townsend}},\
  }\href {\doibase 10.1103/PhysRevLett.102.201301} {\bibfield  {journal}
  {\bibinfo  {journal} {Phys. Rev. Lett.}\ }\textbf {\bibinfo {volume} {102}},\
  \bibinfo {pages} {201301} (\bibinfo {year} {2009}{\natexlab{a}})},\ \Eprint
  {http://arxiv.org/abs/0901.1766} {arXiv:0901.1766 [hep-th]} \BibitemShut
  {NoStop}%
\bibitem [{\citenamefont {Bergshoeff}\ \emph
  {et~al.}(2009{\natexlab{b}})\citenamefont {Bergshoeff}, \citenamefont
  {Hohm},\ and\ \citenamefont {Townsend}}]{Bergshoeff:2009aq}%
  \BibitemOpen
  \bibfield  {author} {\bibinfo {author} {\bibfnamefont {E.~A.}\ \bibnamefont
  {Bergshoeff}}, \bibinfo {author} {\bibfnamefont {O.}~\bibnamefont {Hohm}}, \
  and\ \bibinfo {author} {\bibfnamefont {P.~K.}\ \bibnamefont {Townsend}},\
  }\href {\doibase 10.1103/PhysRevD.79.124042} {\bibfield  {journal} {\bibinfo
  {journal} {Phys. Rev. D}\ }\textbf {\bibinfo {volume} {79}},\ \bibinfo
  {pages} {124042} (\bibinfo {year} {2009}{\natexlab{b}})},\ \Eprint
  {http://arxiv.org/abs/0905.1259} {arXiv:0905.1259 [hep-th]} \BibitemShut
  {NoStop}%
\bibitem [{\citenamefont {Banados}\ \emph {et~al.}(1992)\citenamefont
  {Banados}, \citenamefont {Teitelboim},\ and\ \citenamefont
  {Zanelli}}]{Banados:1992wn}%
  \BibitemOpen
  \bibfield  {author} {\bibinfo {author} {\bibfnamefont {M.}~\bibnamefont
  {Banados}}, \bibinfo {author} {\bibfnamefont {C.}~\bibnamefont {Teitelboim}},
  \ and\ \bibinfo {author} {\bibfnamefont {J.}~\bibnamefont {Zanelli}},\ }\href
  {\doibase 10.1103/PhysRevLett.69.1849} {\bibfield  {journal} {\bibinfo
  {journal} {Phys. Rev. Lett.}\ }\textbf {\bibinfo {volume} {69}},\ \bibinfo
  {pages} {1849} (\bibinfo {year} {1992})},\ \Eprint
  {http://arxiv.org/abs/hep-th/9204099} {hep-th/9204099} \BibitemShut {NoStop}%
\bibitem [{\citenamefont {Oliva}\ \emph {et~al.}(2009)\citenamefont {Oliva},
  \citenamefont {Tempo},\ and\ \citenamefont {Troncoso}}]{Oliva:2009ip}%
  \BibitemOpen
  \bibfield  {author} {\bibinfo {author} {\bibfnamefont {J.}~\bibnamefont
  {Oliva}}, \bibinfo {author} {\bibfnamefont {D.}~\bibnamefont {Tempo}}, \ and\
  \bibinfo {author} {\bibfnamefont {R.}~\bibnamefont {Troncoso}},\ }\href
  {\doibase 10.1088/1126-6708/2009/07/011} {\bibfield  {journal} {\bibinfo
  {journal} {J. High Energy Phys.}\ }\textbf {\bibinfo {volume} {07}},\
  \bibinfo {pages} {011} (\bibinfo {year} {2009})},\ \Eprint
  {http://arxiv.org/abs/0905.1545} {arXiv:0905.1545 [hep-th]} \BibitemShut
  {NoStop}%
\bibitem [{\citenamefont {Giribet}\ \emph {et~al.}(2009)\citenamefont
  {Giribet}, \citenamefont {Oliva}, \citenamefont {Tempo},\ and\ \citenamefont
  {Troncoso}}]{Giribet:2009qz}%
  \BibitemOpen
  \bibfield  {author} {\bibinfo {author} {\bibfnamefont {G.}~\bibnamefont
  {Giribet}}, \bibinfo {author} {\bibfnamefont {J.}~\bibnamefont {Oliva}},
  \bibinfo {author} {\bibfnamefont {D.}~\bibnamefont {Tempo}}, \ and\ \bibinfo
  {author} {\bibfnamefont {R.}~\bibnamefont {Troncoso}},\ }\href {\doibase
  10.1103/PhysRevD.80.124046} {\bibfield  {journal} {\bibinfo  {journal} {Phys.
  Rev. D}\ }\textbf {\bibinfo {volume} {80}},\ \bibinfo {pages} {124046}
  (\bibinfo {year} {2009})},\ \Eprint {http://arxiv.org/abs/0909.2564}
  {arXiv:0909.2564 [hep-th]} \BibitemShut {NoStop}%
\bibitem [{\citenamefont {Kwon}\ \emph {et~al.}(2011)\citenamefont {Kwon},
  \citenamefont {Nam}, \citenamefont {Park},\ and\ \citenamefont
  {Yi}}]{Kwon:2011jz}%
  \BibitemOpen
  \bibfield  {author} {\bibinfo {author} {\bibfnamefont {Y.}~\bibnamefont
  {Kwon}}, \bibinfo {author} {\bibfnamefont {S.}~\bibnamefont {Nam}}, \bibinfo
  {author} {\bibfnamefont {J.-D.}\ \bibnamefont {Park}}, \ and\ \bibinfo
  {author} {\bibfnamefont {S.-H.}\ \bibnamefont {Yi}},\ }\href {\doibase
  10.1007/JHEP11(2011)029} {\bibfield  {journal} {\bibinfo  {journal} {J. High
  Energy Phys.}\ }\textbf {\bibinfo {volume} {11}},\ \bibinfo {pages} {029}
  (\bibinfo {year} {2011})},\ \Eprint {http://arxiv.org/abs/1106.4609}
  {arXiv:1106.4609 [hep-th]} \BibitemShut {NoStop}%
\bibitem [{\citenamefont {Perez}\ \emph {et~al.}(2011)\citenamefont {Perez},
  \citenamefont {Tempo},\ and\ \citenamefont {Troncoso}}]{Perez:2011qp}%
  \BibitemOpen
  \bibfield  {author} {\bibinfo {author} {\bibfnamefont {A.}~\bibnamefont
  {Perez}}, \bibinfo {author} {\bibfnamefont {D.}~\bibnamefont {Tempo}}, \ and\
  \bibinfo {author} {\bibfnamefont {R.}~\bibnamefont {Troncoso}},\ }\href
  {\doibase 10.1007/JHEP07(2011)093} {\bibfield  {journal} {\bibinfo  {journal}
  {J. High Energy Phys.}\ }\textbf {\bibinfo {volume} {07}},\ \bibinfo {pages}
  {093} (\bibinfo {year} {2011})},\ \Eprint {http://arxiv.org/abs/1106.4849}
  {arXiv:1106.4849 [hep-th]} \BibitemShut {NoStop}%
\bibitem [{\citenamefont {'t~Hooft}(1985)}]{tHooft:1984re}%
  \BibitemOpen
  \bibfield  {author} {\bibinfo {author} {\bibfnamefont {G.}~\bibnamefont
  {'t~Hooft}},\ }\href {\doibase 10.1016/0550-3213(85)90418-3} {\bibfield
  {journal} {\bibinfo  {journal} {Nucl. Phys. B}\ }\textbf {\bibinfo {volume}
  {256}},\ \bibinfo {pages} {727} (\bibinfo {year} {1985})}\BibitemShut
  {NoStop}%
\bibitem [{\citenamefont {Mann}\ \emph {et~al.}(1992)\citenamefont {Mann},
  \citenamefont {Tarasov},\ and\ \citenamefont {Zelnikov}}]{Mann:1990fk}%
  \BibitemOpen
  \bibfield  {author} {\bibinfo {author} {\bibfnamefont {R.~B.}\ \bibnamefont
  {Mann}}, \bibinfo {author} {\bibfnamefont {L.}~\bibnamefont {Tarasov}}, \
  and\ \bibinfo {author} {\bibfnamefont {A.}~\bibnamefont {Zelnikov}},\ }\href
  {\doibase 10.1088/0264-9381/9/6/006} {\bibfield  {journal} {\bibinfo
  {journal} {Class. Quant. Grav.}\ }\textbf {\bibinfo {volume} {9}},\ \bibinfo
  {pages} {1487} (\bibinfo {year} {1992})}\BibitemShut {NoStop}%
\bibitem [{\citenamefont {Ghosh}\ and\ \citenamefont
  {Mitra}(1994)}]{Ghosh:1994wb}%
  \BibitemOpen
  \bibfield  {author} {\bibinfo {author} {\bibfnamefont {A.}~\bibnamefont
  {Ghosh}}\ and\ \bibinfo {author} {\bibfnamefont {P.}~\bibnamefont {Mitra}},\
  }\href {\doibase 10.1103/PhysRevLett.73.2521} {\bibfield  {journal} {\bibinfo
   {journal} {Phys. Rev. Lett.}\ }\textbf {\bibinfo {volume} {73}},\ \bibinfo
  {pages} {2521} (\bibinfo {year} {1994})},\ \Eprint
  {http://arxiv.org/abs/hep-th/9406210} {hep-th/9406210} \BibitemShut {NoStop}%
\bibitem [{\citenamefont {Kay}\ and\ \citenamefont {Ortiz}()}]{Kay:2011np}%
  \BibitemOpen
  \bibfield  {author} {\bibinfo {author} {\bibfnamefont {B.~S.}\ \bibnamefont
  {Kay}}\ and\ \bibinfo {author} {\bibfnamefont {L.}~\bibnamefont {Ortiz}},\
  }\href@noop {} {\ }\Eprint {http://arxiv.org/abs/1111.6429} {arXiv:1111.6429
  [hep-th]} \BibitemShut {NoStop}%
\bibitem [{\citenamefont {Ho}\ and\ \citenamefont {Kang}(1998)}]{Ho:1998du}%
  \BibitemOpen
  \bibfield  {author} {\bibinfo {author} {\bibfnamefont {J.-w.}\ \bibnamefont
  {Ho}}\ and\ \bibinfo {author} {\bibfnamefont {G.}~\bibnamefont {Kang}},\
  }\href {\doibase 10.1016/S0370-2693(98)01451-8} {\bibfield  {journal}
  {\bibinfo  {journal} {Phys. Lett. B}\ }\textbf {\bibinfo {volume} {445}},\
  \bibinfo {pages} {27} (\bibinfo {year} {1998})},\ \Eprint
  {http://arxiv.org/abs/gr-qc/9806118} {gr-qc/9806118} \BibitemShut {NoStop}%
\bibitem [{\citenamefont {Liu}\ and\ \citenamefont {Zhao}(2001)}]{Liu:2001qj}%
  \BibitemOpen
  \bibfield  {author} {\bibinfo {author} {\bibfnamefont {W.-B.}\ \bibnamefont
  {Liu}}\ and\ \bibinfo {author} {\bibfnamefont {Z.}~\bibnamefont {Zhao}},\
  }\href {\doibase 10.1088/0256-307X/18/2/353} {\bibfield  {journal} {\bibinfo
  {journal} {Chin. Phys. Lett.}\ }\textbf {\bibinfo {volume} {18}},\ \bibinfo
  {pages} {310} (\bibinfo {year} {2001})}\BibitemShut {NoStop}%
\bibitem [{\citenamefont {Zhou}\ and\ \citenamefont {Liu}(2004)}]{Zhou:2004yq}%
  \BibitemOpen
  \bibfield  {author} {\bibinfo {author} {\bibfnamefont {Z.-A.}\ \bibnamefont
  {Zhou}}\ and\ \bibinfo {author} {\bibfnamefont {W.-B.}\ \bibnamefont {Liu}},\
  }\href {\doibase 10.1142/S0217751X04016416} {\bibfield  {journal} {\bibinfo
  {journal} {Int. J. Mod. Phys. A}\ }\textbf {\bibinfo {volume} {19}},\
  \bibinfo {pages} {3005} (\bibinfo {year} {2004})}\BibitemShut {NoStop}%
\bibitem [{\citenamefont {Hawking}\ \emph {et~al.}(1999)\citenamefont
  {Hawking}, \citenamefont {Hunter},\ and\ \citenamefont
  {Taylor}}]{Hawking:1998kw}%
  \BibitemOpen
  \bibfield  {author} {\bibinfo {author} {\bibfnamefont {S.}~\bibnamefont
  {Hawking}}, \bibinfo {author} {\bibfnamefont {C.}~\bibnamefont {Hunter}}, \
  and\ \bibinfo {author} {\bibfnamefont {M.}~\bibnamefont {Taylor}},\ }\href
  {\doibase 10.1103/PhysRevD.59.064005} {\bibfield  {journal} {\bibinfo
  {journal} {Phys. Rev. D}\ }\textbf {\bibinfo {volume} {59}},\ \bibinfo
  {pages} {064005} (\bibinfo {year} {1999})},\ \Eprint
  {http://arxiv.org/abs/hep-th/9811056} {hep-th/9811056} \BibitemShut {NoStop}%
\end{thebibliography}%


%
%

\end{document}